\documentclass[wssquare]{ws-procs975x65} % for citations in square brackets (consult your editor before picking up this style)
\begin{document}
\title{Muonic atoms and the nuclear structure }

\author{A. Antognini$^*$ for the CREMA collaboration}
%% , 
%% K. Kirch,
%% F. Kottmann,
%% K. Schuhmann and
%% D. Taqqu}

\address{
Institute for Particle Physics, ETH, 8093 Zurich, Switzerland\\
Laboratory for Particle Physics, Paul Scherrer Institute, 5232 Villigen-PSI, Switzerland\\
$^*$E-mail: aldo.antognini@psi.ch\\
}

\begin{abstract}
High-precision laser spectroscopy of atomic energy levels enables the
measurement of nuclear properties. Sensitivity to these properties is
particularly enhanced in muonic atoms which are bound systems of a
muon and a nucleus.  Exemplary is the measurement of the proton charge
radius from muonic hydrogen performed by the CREMA collaboration which
resulted in an order of magnitude more precise charge radius as
extracted from other methods but at a variance of 7 standard
deviations. Here, we summarize the role of muonic atoms for the
extraction of nuclear charge radii, we present the status of the so
called ``proton charge radius puzzle'', and we sketch how muonic atoms
can be used to infer also the magnetic nuclear radii, demonstrating
again an interesting interplay between atomic and particle/nuclear
physics.
\end{abstract}


\bodymatter

\section{What atomic physics can do for nuclear physics}

The theory of the energy levels for few electrons systems, which is
based on bound-state QED, has an exceptional predictive power that
can be systematically improved due to the perturbative nature of the
theory itself~\cite{I.Eides2001, Karshenboim2005}.
On the other side, laser spectroscopy yields spacing between energy
levels in these atomic systems so precisely, that even tiny effects
related with the nuclear structure already influence several
significant digits of these measurements.
Thus, highly accurate atomic transition frequency measurements can be
used as precise and clean probes (purely electromagnetic interaction)
of low energy-QCD properties of the nucleus due to the low energy
nature of the photons articulating the interaction between the
nucleus and the orbiting particle.

A particular class of atoms, called muonic atoms, offer an interesting
opportunity to extract properties of the nucleus with high accuracy.
In these atoms, one or more electrons are substituted by a muon, which
is a fundamental particle having the same electromagnetic properties
as the electron but with a much larger mass ($m_\mu\approx200 m_e$).
For example muonic hydrogen ($\mu p$) is the bound system of a
negative muon and a proton, muonic helium ion ($\mu$He$^+$) a muon
bound to an alpha particle.
The atomic properties are strongly affected by the orbiting particle
mass $m$, e.g., the Bohr energy scales linearly with $m$ while the
Bohr radius as $1/m$, resulting already for low-Z atoms in muonic
binding energies of several keV and in a so strong overlap of the muon
wave functions with the nucleus that the energy levels are
considerably (\%-level) affected by the nucleus finite size.
A paradigmatic example is $\mu$p whose laser spectroscopy yielded a
very precise determination of the proton charge radius~\cite{Pohl2010,
  Antognini2013a}.

\section{Charge and magnetic radii of the proton from scattering}

The scattering process between charged particles without internal
structure, as for example electron-electron scattering can be fully
described within QED.
Oppositely, when describing electron-proton scattering, form factors
need to be introduced to parameterize the complexity of the nuclear
structure.
They contain dynamical information on the electric and magnetic
currents in the nucleus defining the response to the electromagnetic
fields.
As a consequence of current conservation and relativistic invariance,
for the spin-1/2 nuclei, as protons, only two form factors are required.
Experimentally these form factors can be accessed through 
measurements of the elastic differential cross section which
in the one-photon approximation is~\cite{Punjabi2015a} %Perdrisat2007,
\begin{equation}
\Big(\frac{d\sigma}{d\Omega}\Big)_\mathrm{elastic}=\Big(\frac{d\sigma}{d\Omega}\Big)_\mathrm{Mott} \times \frac{1}{1+\tau}\Big( G^2_E(Q^2)+\frac{\tau}{\varepsilon}G^2_M(Q^2)  \Big) ,
\label{eq:Mott}
\end{equation}
where the Mott cross section applies for point-like particles and is
fully calculated in the QED framework.
$G_E(Q^2)$ and $G_M(Q^2)$ are the electric and magnetic Sachs form
factors, while $\tau=Q^2/4M^2$ and $\epsilon^{-1}=1+2(1+\tau)
\tan^2{(\theta/2)}$ are kinematical variables with $\theta$ being the
electron scattering angle and $M$ the nucleus mass.
At $Q^2=0$ the form factors correspond to the total charge in units of
$e$ and magnetic moment in units of the proton magneton: for the
proton $G^p_E(0)=1$ and $G^p_M(0)=2.793$. 
So in first approximation at low momentum exchange the response of the
nucleus to electromagnetic fields is ruled by its charge and magnetic
moment.

Viewed as a Taylor series the charge and the magnetic moment are the
first terms in an infinite list of parameters which describes the
interaction of the proton with the electromagnetic 
fields~\cite{Epstein2014}.
The next parameters would be the slopes of the electric and magnetic
form factors at zero momentum exchange:
\begin{equation}
R_E=-\frac{6}{G_E(0)}\frac{dG_E}{dQ^2}\Big|_{Q^2=0} \qquad \qquad \mbox{and} \qquad \qquad
R_M=-\frac{6}{G_M(0)}\frac{dG_M}{dQ^2}\Big|_{Q^2=0} \,.
\label{eq:radii}
\end{equation}
These equations represent the covariant definition of charge and
magnetic radii, which in a non-relativistic approximation correspond to
the second moments of the electric charge and magnetization
distributions $\rho_{E}$ and $\rho_M$ of the nucleus
\begin{equation}
R_{E/M}^2\approx\int d\vec{r}\, \rho_{E/M}({\vec r})r^2.
\end{equation}
Any hadron/nuclear theory must reproduce these radii being parameters as
fundamental as the charge, mass and magnetic moment.
Although lattice QCD shows  an impressive
progress~\cite{Green2014}, currently these radii can not be
accurately predicted from ab-initio theories and their knowledge
relies on experiments~\cite{Punjabi2015a, Bernauer2014}. %Perdrisat2007, Zhan2011

The traditional way to extract the form factors from the measured
differential cross sections is based on the Rosenbluth separation
techniques which consist in plotting the reduced cross section
$\sigma_\mathrm{red}$ versus $\varepsilon$:
\begin{equation}
\sigma_\mathrm{red}\equiv \frac{\varepsilon (1+\tau)}{\tau}
\; \frac{\Big(\frac{d\sigma}{d\Omega}\Big)_\mathrm{elastic}}{\Big(\frac{d\sigma}{d\Omega}\Big)_\mathrm{Mott}}= G_M^2+ \frac{\varepsilon}{\tau} G_E^2 \,.
\label{eq:red}
\end{equation}
The reduced cross section is  linear
in $\varepsilon$, with the slope proportional to $G_E^2$ and
the intercept equal to $G_M^2$.
So both form factors can be deduced by measuring $\big(\frac{d\sigma}{d\Omega}\big)_\mathrm{elastic}$
at several values of $\varepsilon$ which is achieved by
varying the electron beam energy and the electron scattering angle
while keeping $Q^2$ fixed.

After this $G_E/G_M$ separation, each measured form factor can be fitted
with a polynomial expansion of the form~\cite{Sick2003}
\begin{equation}
 G_{E/M}(Q)= G_{E/M}(0) \Big[ 1-\frac{Q^2}{6}\langle r_{E/M}^2\rangle +\frac{Q^4}{120}\langle r_{E/M}^4\rangle-\dots \Big] \; ,
\label{eq:taylor}
\end{equation}
where $\langle r_{E/M}^N\rangle$ represent the $N$-th moments of the charge/magnetic
distributions ($\langle r_{E/M}^2\rangle=R_{E/M}^2$).
At very low $Q^2$, one could hope that the higher moments terms are
sufficiently small, such that the $\langle r_{E/M}^2 \rangle$-term can be
determined without using a specific model for the form factor.
However, at low $Q^2$ also the $\langle r_{E/M}^2 \rangle$-term
becomes increasingly small relative to the first term of the expansion
resulting in a loss of sensitivity.
So in practice to fit the measured form factors and extract the radii
it is necessary to include data at intermediate $Q^2$.
As cross sections data are available only down to a minimal $Q^2$, and
because an extrapolation to $Q^2=0$ is required, the choice of
the fit function (form factor model) is very important.

This extrapolation is even more challenging for the magnetic radii
because of the $\varepsilon/\tau$-dependence in Eq.~(\ref{eq:Mott})
which results in an additional suppression of sensitivity (at low
$Q^2$) of the measured cross sections to $G_M$ compared to $G_E$.
Consequently, the increased uncertainties of $G_M$ at low $Q^2$ yields 
magnetic radii with larger uncertainties relative to charge radii.
This calls for alternative determinations of the magnetic radii such
as from polarized-recoil scattering~\cite{Punjabi2015a} or atomic
spectroscopy.

%%%%%%%%%%%%%%%%%%%%%%%%%%%%%%%%%%%%%%%%%%%%%%%%%%%%%%%%%%%%%%%%%%%%%%%%%%%%%%%%%%%%
\section{Charge and magnetic  radii of the proton from atomic physics}

The finite radius of the nucleus implies that its charge  is
smeared over a finite volume.
For hydrogen-like S-states there is a non-negligible probability that the ``orbiting'' particle
is spending some time inside the nuclear charge distribution, thus
experiencing a reduced electrostatic attraction as compared to a
point-like nucleus.
This reduced attraction caused by the modification of the Coulomb
potential for very small distances is giving rise to a shift of
the atomic energy levels which for S-states H-like systems  in leading order
reads~\cite{I.Eides2001, Karshenboim2005}
\begin{equation}
\Delta E_\mathrm{finite \;size}= \frac{2\pi Z\alpha}{3} |\phi^2(0)|^2 R_E^2
                      =  \frac{2 m_r^3 (Z\alpha)^4}{3n^3} R_E^2 \;,
\label{eq:finite-size}
\end{equation}
where $\phi(0)$ is the wave function at the origin in coordinate space,
$m_r=mM/(m+M)$ the reduced mass of the atomic system with $m$ being
the orbiting particle mass, and $M$ the nucleus mass, $\alpha$ the
fine structure constant, $Z$ the charge number of the nucleus and $n$
the principal quantum number.

The $m_r^3$ dependence of Eq.~(\ref{eq:finite-size}) reveals the
advantages related with muonic atoms.
As the muon mass is 200 times larger than the electron mass, the
muonic wave function strongly overlaps with the nucleus ensuing a large
shift of the energy levels due to the nuclear finite size.
Thus, the muonic bound-states represent ideal systems for the precise
determination of nuclear charge radii $R_E$~\cite{Pohl2010, Antognini2013a, Pohl2013}. %~\cite{Pachucki1999, Borie2012}.

Because of this sensitivity to the finite size a moderate (20 ppm)
accuracy in the measurement of the 2S-2P transition in $\mu$p is
sufficient to extract the proton charge radius very accurately
($5\times 10^{-4}$ relative accuracy)~\cite{Pohl2010, Antognini2013a}.
In regular H, the accuracies of the transition frequency
measurements, also relative to the line-widths, have to
be much higher (see Table~\ref{tab:1}) to compete with this value.
By combining in a least-square adjustment all high-precision frequency
measurements in H available to date, as accomplished by the CODATA
group, a proton charge radius with an accuracy of about 1\% is
obtained.

Atomic spectroscopy can be used also to extract magnetic radii.
This is achieved through precision measurement of hyperfine
splittings~\cite{Volotka2005, Dupays2003, Antognini2013a}.
For H-like systems, in leading approximation the HFS is given by the
magnetic interaction between the nucleus $\vec{\mu}_N$ and the
orbiting particle $\vec{\mu}_m$ magnetic moments, described by~\cite{I.Eides2001, Karshenboim2005}
\begin{equation}
H \sim\vec{\mu}_N\cdot\vec{\mu}_m\; \delta(\vec{r})\; ,
\end{equation}
which results in an energy splitting of the 1S state given by
the Fermi energy
\begin{equation}
E_F=\frac{8}{3} \frac{Z^3\alpha^4 m_r^3}{mMn^3}  \mu_N  \;.% \frac{m^3m_p^3}{(m+m_p)^3} 
\label{eq:Fermi}
\end{equation}
The finite size correction to this splitting, which is of second
order in perturbation theory, is~\cite{Dupays2003, Karshenboim2005}
\begin{equation}
\Delta E_\mathrm{Zemach}=-2 (Z\alpha) m_r  \, E_F\; R_Z
\label{eq:Zemach-contribution}
\end{equation}
where the Zemach radius $ R_Z$
is defined as an integral of the charge and magnetic form factors 
\begin{equation}
R_Z=-\frac{4}{\pi}\int_0^\infty \frac{dQ}{Q^2} \; \Big( G_E(Q^2)
\frac{G_M(Q^2)}{1+\kappa_p}-1 \Big) \; ,
\label{eq:def-Rz-1}
\end{equation}
(with $\kappa_p$ the proton anomalous magnetic moment).
In a non-relativistic approximation $R_Z$ can be expressed, by the
first moment of the convolution between charge and magnetic
distributions $\rho_E(r)$ and $\rho_M(r)$ in coordinate space
\begin{equation}
R_Z=\int d^3 {\bf r} \; |{\bf r}| \int d^3{\bf r}' \rho_E(\mathbf{r}-\mathbf{r'}) \rho_M(\mathbf{r'}).
\label{eq:def-Rz-2}
\end{equation}
When assuming form factor models or using measured form factor data,
the magnetic radius can be extracted from the Zemach radius.
Thus, accurate measurements of the HFS in $\mu$p and H can be used as
complementary ways to obtain a precise value of the proton magnetic
radius, or alternatively, as presented in~\cite{Karshenboim2014_rZ} a
self-consistent value of $R_E^2+R_M^2$.
%

%%%%%%%%%%%%%%%%%%%%%%%%%%%%%%%%%%%%%%%%%%%%%%%%%%%%%%%%%%%%%%%%%%%%%%%%%%%%%%%%%%%%%%%%%%%%%%%%%%%%%%
\section{The proton charge  radius puzzle}
\label{sec-1a}

Three complementary routes to the proton charge radius have been
undertaken: the historical method relies on elastic 
electron-proton scattering, the second one on high-precision laser
spectroscopy in H, and the third one on high sensitivity
laser spectroscopy in $\mu$p.
The value extracted from $\mu$p~\cite{Pohl2010, Antognini2013a} with a
relative accuracy of $5\times 10^{-4}$ is an order of magnitude more
accurate than obtained from the other methods.
Yet the value is 4\% smaller than derived from electron-proton
scattering~\cite{Sick2014b, Bernauer2014, Lee2015} and H
spectroscopy~\cite{DeBeauvoir2000} with a disagreement at the $7\sigma$
level.
\begin{figure}[h]
\begin{center}
\vspace{-0.95mm}
\includegraphics[width=3.4in]{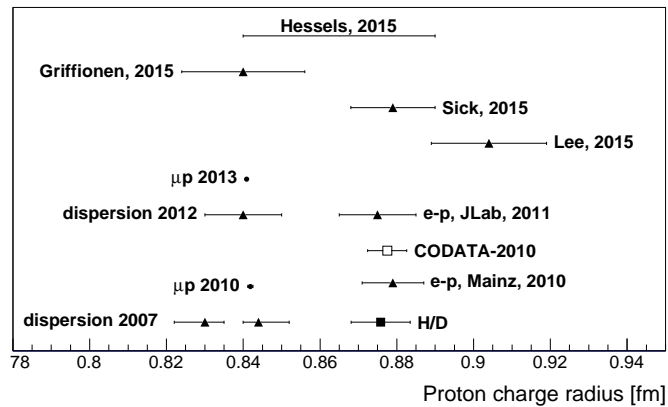}
\end{center}
\vspace{-0.95mm}
\caption{Proton  charge radii
   determined from spectroscopy of muonic atoms  (full circles), from electron
   scattering (triangles) and from H/D spectroscopy (full squares). }
\label{fig:radii}
\end{figure}

The most recent evaluations of the proton charge radius are summarized
in Fig.~\ref{fig:radii}.
The most precise values are extracted from two transition frequency
measurements in $\mu$p.
By combining them we obtained a 2S-2P$_{1/2}$ splitting
of
%
%\begin{equation}
$\Delta E_{2S-2P_{1/2}}^\mathrm{exp}=202.3706(23)~\mathrm{ meV}$ %=  48932.99(55)~\mathrm{GHz}$
%\end{equation}
%
equivalent to a frequency of  48932.99(55)~GHz,  limited by statistics while the systematic effects are at the 300~MHz
level~\cite{Antognini2013a}.  Equating this experimental value with the theoretical prediction %summarized
%in~\cite{Antognini2013}
%
%
\begin{equation}
E_L^\mathrm{th}=206.0336(15)~\mathrm{[meV]}-5.2275(10)~\mathrm{ \Big[\frac{meV}{fm^2}\Big]} R_E^2+0.0332(20)~\mathrm{[meV]}
\label{eq:lamb-shift-mup}
\end{equation}
yields the proton charge radius $R_E$  in fm.
The first term of Eq.~(\ref{eq:lamb-shift-mup}) accounts for QED
contributions, the second one for finite size effects, and the third
one for the two-photon exchange (TPE) contribution which is a
second-order perturbation theory contribution related with the proton
structure. %% which can not be
%% simply calculated in the framework of bound-state QED.
%
In the last years as summarized in~\cite{Pohl2013, Karshenboim2015a,
  Peset2015} various cross checks and refinements of bound-state QED
and TPE %~\cite{Alarcon2014, Birse2012} 
calculations needed for the
extraction of $R_E$ from $\mu$p have been performed, but no
substantial missing effects have been found that could explain the
discrepancy.

The typical systematics affecting the atomic energy levels are
substantially suppressed in $\mu$p due to the stronger binding.
The internal fields and the level separation of the muonic atoms are
greatly enhanced compared to regular atoms making them insensitive to
external fields (AC and DC Stark, Zeeman, black-body and
pressure shifts).
Thus $\mu$p turns out to be very sensitive to the proton charge radius
($m_r^3$-dependence) and insensitive to systematics which typically  scale
as $\sim1/m_r$.
%% %

Special attention was devoted to the analysis of electron-proton
scattering data and the issues related with the extrapolation procedure.
Starting from fit functions given by truncated general series
expansions such as Taylor, splines and polynomials a large progress
has been achieved in the last years by the use of various techniques:
enforcing analyticity~\cite{Epstein2014, Lee2015}, constraining the low
$Q^2$ behavior of the form factor assuming a large-$r$ behavior of the
charge distribution~\cite{Sick2014b} or by using proton
models~\cite{Lorenz2015}.
Tension exists between various electron-proton data analysis:
some give results compatible with $\mu$p~\cite{Lorenz2015, Griffioen2015, Higinbotham2015}, some at variance~\cite{Sick2014b, Bernauer2014, Lee2015, Distler2015}.
%% %
Because data at even lower $Q^2$ would facilitate the extrapolation
at $Q^2=0$, two electron-proton experiments have been initiated, one
at JLAB~\cite{Gasparian2014}, the other one at MAMI
Mainz~\cite{Mihovilovic2014}.
A comparison between muon-proton and electron-proton scattering within
the same setup as proposed by the MUSE~\cite{Gilman2013} collaboration
at PSI could disclose a possible violation of muon-electron
universality.

Several ``beyond standard model'' BSM extensions have been studied but
the majority of them have difficulties to resolve the discrepancy
without conflicting with other low energy constraints.
Still some BSM theories can be formulated but they require fine-tuning
(e.g. cancellation between axial and vector components), targeted
coupling (e.g. preferentially to muons) % or to muon-proton)
and are
problematic to be merged in a gauge invariant way into the standard
model~\cite{Karshenboim2014, Carlson2015}.
Breakdown of the perturbative approach in the electron-proton
interaction at short distances, as well as the interaction with sea
$\mu^+\mu^-$ and $e^+e^-$ pairs and unusual proton structure have been
suggested as possible explanation but without conclusive
quantification~\cite{Jentschura2015} .

Summarizing, currently the discrepancy persists even though recent
reanalysis of scattering data have led to larger uncertainties of the
extracted proton radius.
New data from muonic deuterium and helium, from H spectroscopy and
electron-proton scattering holds the potential to clarify the
situation in the near future.

%%%%%%%%%%%%%%%%%%%%%%%%%%%%%%%%%%%%%%%%%%%%%%%%%%%%%%%%%%%%%%%%%%%%%%%%%%%%%%%%%%%%%%%%%%%%%%%%%%%%%%
\section{The proton radius from H spectroscopy}
\label{sec-1c}

In a simplified way, the hydrogen S-state energy levels can be
described by
\begin{equation}
  E(nS)=\frac{R_\infty}{n^2}+\frac{L_{1S}}{n^3} \; ,
\label{eq:H}
\end{equation}
where $R_\infty=3.289 \,841 \,960 \,355(19)
\times 10^{15} $~Hz is the Rydberg constant and 

\begin{equation}
L_{1S}\simeq 8171.636(4)~\mathrm{[MHz]} + 1.5645~\mathrm{\Big[\frac{MHz}{fm^2}\Big]} R_E^2\;
\label{eq:lamb-shift-H}
\end{equation}
the 1S Lamb shift given by bound-state QED contributions.
The different $n$-dependence of the two terms in Eq.~(\ref{eq:H}) permits
to extract both $R_\infty$ and $L_{1S}$ (thus $R_E$ ) from at least two
frequency measurements in H.

Being the most precisely known transition (relative accuracy of $4\times
10^{-15}$)~\cite{Matveev2013} and having the 
largest sensitivity to $R_E$, usually the
1S-2S transition is used.
By combining it with a second transition 
measurement, $R_\infty$ is eliminated and $R_E$ can be extracted.
%as shown in Fig.~\ref{}.
%
When taken individually, the various $R_E$ values extracted from H
spectroscopy by combining two frequency measurements (2S-4S, 2S-12D,
2S-6S, 2S-6D, 2S-8S, 1S-3S as ``second'' transition~\cite{DeBeauvoir2000}) are statistically
compatible with the value from $\mu$p.
Only the value extracted by pairing the 1S-2S and the
2S-8D transitions is showing a $3\,\sigma$ 
deviation while all the others differ only by $\lesssim 1.5\,\sigma$.

So the $4\,\sigma$ discrepancy between the proton charge radius from
$\mu$p and H spectroscopy emerges only after an averaging process
(mean square adjustments of all measured transitions) of the various
``individual'' determinations and consequently is less startling than
it looks at first glance.
A small systematic effect common to the H measurements could be
sufficient to explain the deviation between $\mu$p and H results.
This fact becomes even more evident if we consider the frequency
shifts (absolute and normalized to the line-width) necessary to match
the $R_E$ values from $\mu$p and H, as summarized for selected
transitions in Table~\ref{tab:1}.
\begin{table}[tbh]
\tbl{Relative accuracy of the various transition measurements in H,
  and hypothetical shift of the measured transition frequencies
  needed to match the $R_E$ from H and $\mu$p. This shift is expressed also relative
  to the experimental accuracy $\sigma$, and to the transition
  effective line-widths $\Gamma_\mathrm{eff} $. }  {
                  \begin{tabular}{l||l|rrr}
                    \hline 
                    Transition       & Relative accuracy & $\quad$ Shift in $\sigma$  & $\quad$ Shift in Hz & $\quad$ Shift in line-width\\

                    \hline 
%                    \multicolumn{4}{l}{Explain the discrepancy by shifting the}\\
%                    \hline 
                    $\mu$p(2S-2P) & $2\times 10^{-5}$   &  $100 \, \sigma $    & 75 GHz    & $4\, \Gamma_\mathrm{eff}$\\     
                    H(1S-2S)      & $4\times 10^{-15}$  &  $4'000 \, \sigma $  & 40 kHz    & 40$\, \Gamma_\mathrm{eff}$\\     
                    H(2S-4P)      & $3\times 10^{-11}$  &  $1.5\, \sigma    $  &  9 kHz    & $7\times 10^{-4} \,\Gamma_\mathrm{eff}$\\     
                    H(2S-2P)      & $1\times 10^{-6}$  &  $1.5\, \sigma $     &  5 kHz    &  $ 7\times 10^{-4} \,\Gamma_\mathrm{eff}$\\     
                    H(2S-8D)      & $9\times 10^{-12}$  &  $ 3 \, \sigma $     &  20 kHz   & $ 2\times 10^{-2} \,\Gamma_\mathrm{eff}$\\     
                    H(2S-12D)     & $1\times 10^{-11}$  &  $ 1 \, \sigma $     &  8 kHz    &  $ 5\times 10^{-3} \,\Gamma_\mathrm{eff}$\\     
                    H(1S-3S)      & $4\times 10^{-12}$  &  $ 1 \, \sigma $     &  13 kHz   & $ 5\times 10^{-3} \,\Gamma_\mathrm{eff}$\\     
                    \hline 
                   \end{tabular}
}
\label{tab:1}
\end{table}
Obviously the discrepancy can not be solved by slightly tuning
(shifting) the measured values of the 1S-2S transition in H and the
2S-2P transitions in $\mu$p because it would require displacements
corresponding to $4000\,\sigma$ and $100\,\sigma$, respectively.
Expressing the required frequency shift relative to the
line-width as in the last column allows to
better recognize some aspects of the experimental challenges. 
For example a  shift of only $7\times 10^{-4} \,\Gamma$ of the
2S-4P transition would be sufficient to explain the
discrepancy.
A control of the systematics  which could distort and shift the
line shape on this level of accuracy is far from being a trivial task.
Well investigated are the large line broadening owing to
inhomogeneous light shifts which results in profiles with effective
widths much larger than the natural
line-widths~\cite{DeBeauvoir2000}.

Another exemplary correction relevant in this context, named quantum
interference, has been brought recently back to
attention~\cite{Horbatsch2010}, and has lead to various reevaluations
of precision experiments. %~\cite{Amaro2015}.
An atomic transition can be shifted by the presence of a neighboring
line, and this energy shift $\delta E $, as a rule of thumb, amounts maximally
to
%
%\begin{equation}
$\frac{\delta E }{ \Gamma} \approx \frac{\Gamma }{D} $
%\end{equation}
%
where $D$ is the energy difference between the two resonances
and $\Gamma$ the transition line-width.
Thus, if a transition frequency is aimed with an absolute accuracy of
$\Gamma/x$, then the influence of the neighboring lines with $D \le
x\Gamma$ has to be considered.
The precise evaluation of these quantum interference effects are
challenging because they require solving numerous differential
equations describing the amplitude of the total excitation and
detection processes from initial to final state distributions which
depends on the details of the experimental setup.

Generally speaking, transition frequencies involving states with large
$n$ are more sensitive to systematic effects caused
by external fields. 
Emblematic is the $n^7$-dependence of the  Stark effect.
Motivated by the possibility that minor effects in H could be
responsible for the discrepancy, various activities have been
initiated in this field: at MPQ Garching the 2S-4P~\cite{Beyer2015} and 1S-3S
transitions are addressed, at LKB Paris the 1S-3S~\cite{Galtier2015}, and at the
Toronto university   the 2S-2P~\cite{Vutha2012}.

The ``second'' (beside the 1S-2S transition) transition frequency
measurement in H can be interpreted as a $R_\infty$ determination.
Optical spectroscopy of H-like ions between circular Rydberg states
where the nuclear size corrections are basically absent, the QED
contributions small, and the line-widths narrow can be used as
alternative determination of $R_\infty$~\cite{Tan2011}.
Another way to $R_\infty$ is through spectroscopy of muonium and
positronium atoms which are purely leptonic systems where
uncertainties related with the finite size are absent~\cite{Cooke2015}.

\section{Hyperfine splitting in $\mu$p and $\mu^3$He$^+$ }

As a next step, we plan to prepare the measurement by means of laser
spectroscopy of the ground state hyperfine splitting (1S-HFS) in
$\mu$p and $\mu^3$He$^+$ with few ppm relative accuracy.
Similar activities in $\mu$p exist at RIKEN-RAL and
J-PARC~\cite{Adamczak2012, Sato2014}.
The theoretical prediction for the 1S-HFS in $\mu$p is
approximately~\cite{Volotka2005, Dupays2003, Faustov2014,
  Carlson2011a}
\begin{equation}
\Delta E_\mathrm{HFS}^\mathrm{th}=182.819(1)~\mathrm{[meV]}-1.301\,\mathrm{ \Big[\frac{meV}{fm}\Big]} R_Z+ 0.064(21)~\mathrm{[meV]} \;,
\label{eq:HFS-mup}
\end{equation}
where the first term includes the Fermi energy, QED corrections,
hadronic vacuum polarization, recoil corrections and weak
interactions.
These contributions are known well enough.
The second term is the finite size contribution, which is proportional
to $R_Z$.
It contains also some higher order mixed radiative finite-size
corrections.
The third term is given by the proton polarizability contribution.

By comparing the theoretical prediction with the experiment, it will
become possible to deduce $R_Z$ with a relative accuracy better than
$5\times 10^{-3}$ provided that the polarizability contribution will
be improved below 10\% relative accuracy.
This contribution can be computed using a dispersive approach and
measured proton polarized structure function $g_1$ and
$g_2$~\cite{Faustov2014, Carlson2011a}
%~\cite{Slifer, Sulkosky:2013gra}, 
or via chiral perturbation theories (ChPT) \cite{Hagelstein2016a}. 
An improvement of this contributions is conceivable in the near future
due to the considerable advance in ChPT \cite{Hagelstein2016} and due to various ongoing
measurements of the proton structure functions at JLAB using polarized
target and beams.

For $\mu^3$He$^+$ the situation is conceptually similar to $\mu$p.
The theoretical predictions assumes the same form as in
Eq.~(\ref{eq:HFS-mup}) but with different numerical values. 

The motivations for these experiments are several:

\begin{itemize}
\item{\bf Bound-state QED in H and understanding of the 21 cm line}
  
The uncertainty of $R_Z$ presently limits, together with the
polarizability contribution, the theoretical prediction of the 1S-HFS
in H.  Therefore, the comparison between the experimental 1S-HFS
value in H, which has a relative accuracy smaller than
$10^{-12}$, with the theoretical predictions is
limited by the uncertainty of the proton structure contributions. This
situation can be improved by complementary measurements in $\mu$p %% which
%% yield an improved determination of the Zemach
%% radius~\cite{Karshenboim2005, Dupays2003}
opening the way for a  test of the HFS in H
at the $10^{-7}$ level of accuracy. %\\[-2mm]

\item{\bf Understanding of the proton structure}

Practically, from the Zemach radius the magnetic radius can be
obtained by using  form factor models or measured form factor data.
As the determination of $R_M$  from elastic
electron-proton scattering is very challenging due to the loss of
sensitivity for the magnetic form factor with decreasing momentum
exchange, a precise measurement of $R_Z$   from the muonic HFS
represents a valuable complementary route to $R_M$.
It can be used also to sort out a 8\% discrepancy between
$R_M$ as extracted from the recent
unpolarized electron-proton cross sections measurements in
Mainz, and as deduced from polarized-recoil data
at JLAB~\cite{Epstein2014, Bernauer2014, Lee2015, Lorenz2015}.

Currently, we cannot determine the radii and the form factors
accurately from theory, although lattice QCD is making impressive
progress on this issue~\cite{Green2014}.
A precise measurement of $R_Z$ from $\mu$p and its comparison
with correlative measurements from scattering experiments bears the
potential to push the frontier of our understanding of the complex
non-perturbative nature of the proton structure which has been
deeply reviewed in the last 15 years especially due to polarization
data and the development of theoretical tools such as chiral
perturbation theory.  %~\cite{Bernard2013}.

The interplay between the muonic measurement and investigations of
the proton structure can be articulated in several ways. 
As mentioned previously $R_Z$ ($R_M$) represents a benchmark for the
understanding of the proton structure.
Extraction of a precise value for the Zemach radius from the $\mu$p
1S-HFS measurement requires the knowledge of the proton polarizability
contribution which requires modeling of the proton and data from
scattering (ChPT, $g_1$ and $g_2$ structure functions).
Inverting this logic, a precise value of $R_Z$ from scattering
data~\cite{Distler2011} can be thus used, when paired with the $\mu$p
HFS, to check the polarizability contribution.

\item{\bf Nuclear physics from $\mu^3$He$^+$}
  
  Nuclei like $^3$He are calculable very precisely by a wide variety
  of ab-initio methods and so provide an important comparison between
  experiment and theoretical models of both the nuclear interactions
  (potential) and the electromagnetic currents~\cite{Sick2014c}.  The magnetic
  distribution and magnetic radii turn out to be very sensitive to
  the meson-exchange currents.
  A very hot topic in hadronic physics is to
  measure various parton distributions  to see the
  quark spin distribution within protons and neutrons.  The same
  should be done for nucleon spin distributions in
  nuclei. 

\end{itemize}

\section{Conclusions}

Precision measurements in muonic atoms have triggered a plethora of
theoretical works and experimental investigations in various fields of
physics showing the potential and interdisciplinarity  of these
precision experiments~\cite{Carlson2015}.
Spectroscopy of the 2S-2P splittings in $\mu$p, $\mu$d, $\mu^3$He$^+$
and $\mu^4$He$^+$ has been accomplished by the CREMA collaboration.
Besides the proton charge radius, soon new accurate values of the
deuteron and $^3$He and $^4$He nuclear radii will be extracted from
these measurements providing insights into the proton radius puzzle,
and benchmarks to check few-nucleon ab-initio calculations.
%% %
Moreover they can be used as anchor point for the $^6$He-$^4$He and
$^8$He-$^4$He isotopic shift measurements~\cite{Lu2013} and their
knowledge opens the way to enhanced bound-state QED
tests for one- and two-electrons systems in
``regular'' He$^+$~\cite{Herrmann2009} and He~\cite{Kandula2011}.

Spectroscopy of HFS transitions in $\mu$p and $\mu^3$He$^+$ provides a
natural continuation of the CREMA program. % aiming at the nuclear
%magnetic radii which are of difficult access from scattering.
%
Letting aside the proton radius puzzle related ``new physics''
searches %~\cite{Keung2015, Karshenboim2014, Carlson2015b, Brax2014},
the 1S-HFS in $\mu$p and $\mu^3$He$^+$ measurements impact three
aspects of fundamental physics: bound-state QED in H-like systems, our
understanding of the magnetic distributions and the low-energy spin
structure of proton and $^3$He nucleus.

\section*{Acknowledgments}
This work is supported by the Swiss National Science Foundation
Projects No. 200021L\_138175 and No.  200020\_159755. We acknowledge
fruitful discussions with N.~Barnea, R. Wiringa, I. Sick and V. Pascalutsa.

\vspace{-1mm}

%
%\begin{verbatim}
%\bibliographystyle{ws-procs975x65}
%\bibliography{library.bib}
%\end{verbatim}

\end{document}